\documentclass[a4paper,11pt]{article}

\usepackage{amsmath}
\usepackage{amssymb}
\usepackage{amsthm}
\usepackage{graphicx}
\usepackage{enumerate}
\usepackage{url}
\usepackage{authblk}

\usepackage[a4paper, total={6in, 8in}]{geometry}

\theoremstyle{definition}

\begin{document}

\title{Artificial Intelligence in Music and Performance: A Subjective Art-Research Inquiry}
\author[1]{Baptiste Caramiaux}
\author[2]{Marco Donnarumma}
\affil[1]{Universit\'e Paris-Saclay, CNRS, Inria, LRI}
\affil[2]{Independent Researcher}
\date{}                     
\setcounter{Maxaffil}{0}
\renewcommand\Affilfont{\small}

\maketitle

\begin{abstract}

This article presents a five-year collaboration situated at the intersection of Art practice and Scientific research in Human-Computer Interaction (HCI). At the core of our collaborative work is a hybrid, Art and Science methodology that combines computational learning technology -- Machine Learning (ML) and Artificial Intelligence (AI) -- with interactive music performance and choreography. This article first exposes our thoughts on combining art, science, movement and sound research. We then describe two of our artistic works \textit{Corpus Nil} and \textit{Humane Methods} -- created five years apart from each other -- that crystallize our collaborative research process. We present the scientific and artistic motivations, framed through our research interests and cultural environment of the time.
We conclude by reflecting on the methodology we developed during the collaboration and on the conceptual shift of computational learning technologies, from ML to AI, and its impact on Music performance.

\end{abstract}

\section{Introduction}

In many contemporary societies the pervasiveness of technology is constantly expanding. From communication to social networks, digital health and welfare services, every aspect of social life in industrialized societies is being captured by technology with the objective of human enhancement, optimized services, or automated management. Among these technologies, Machine Learning (ML) and the broader field of Artificial Intelligence (AI) received considerable attention in the past decades. In this chapter, we use an autoethnographic approach to present and discuss the hybrid methodology that we developed in five years of collaborative research across computation, science and art. The analysis we offer here combines insight from academic research in Human-Computer Interaction (HCI), in particular body-based interaction, and from aesthetic research in the performing arts to inspect and question the role of ML and AI in our practices.

The past decade has shown that data-driven learning-based algorithms can succeed in many tasks that were unthinkable not too long ago. Since the deep learning breakthrough in 2012 \cite{krizhevsky2012imagenet}, these algorithms have been shown to recognize images as well as humans do, to acquire motor control capacities from few observations, and to understand and respond to several kinds of human languages. These breakthroughs prompted major investments in the field, thus accelerating technological advance at an exponential pace (the number of attendees at major conferences or the number of published papers at online repository like \textit{arxiv} show the radical increase of interest in the topic). A consequence of the impressive advance of the field is a frantic race to embed ML-based algorithms in most digital services, without offering ways for users to acquire a basic literacy of the technology. The general paradigm behind the development of these technologies is based on centralized ownership by tech companies or institutions. By interacting with these technologies, people produce large amount of usage data, which, in turn, is used to train the underlying algorithms. This type of technology is therefore detached from people’s control and understanding.

Artists have historically been among the first to question  technological innovations (see, for instance, the organization Experiments in Art and Technology, E.A.T., founded in 1967). They have often been early adopters and disruptors of new technological tools, and ML is not an exception \cite{caramiaux2019ai}. On one hand, many artists use ML to enrich the way they work with their preferred media. On the other, some artists use ML, and in particular AI, to shed light onto certain facets of these same tools  which can be invisible or taken for granted by public opinion, media, or institutions.

Our own path, which we elaborate next, has taken different directions throughout the past five years of collaboration. Initially, our approach was to use ML as a tool to design improvised and markedly physical musical performance, exploring the relation between computation and corporeality. Eventually, our approach shifted to utilize and reflect on AI as an actor in a performance, an entity whose functioning can be used to question the understanding of computational intelligence in Western society.
As a team doing research in overlapping fields, we tried to combine our respective methodologies in a zone of encounter, where \emph{practice-based research} meets \emph{artistic intervention in research}. The former  involves the use of artistic practice as a means of research. The latter entails that a creative, artistic process acts upon scientific research objectives. Naturally, such a hybrid methodology has been explored by others before  us~\cite{fdili2019making,edmonds2010relating}.

In this chapter, we intend to extract the subjective aspects of our collaborative works and to discuss the methodological perspective that our particular mode of collaboration offers to the study of ML and AI, as well as to the study of their impact on HCI and the performing arts. The chapter is structured as follows. First, we provide general thoughts on combining scientific and artistic practices; these observations will help us give context to the methodology of our collaborative work. The next two sections discuss the conception and development of our collaborative artworks, two hybrid performances of computational music and choreography, \emph{Corpus Nil} and \emph{Humane Methods}.
We present the scientific and artistic drives behind each project, highlighting scientific, artistic and cultural contexts. Finally, we provide a closing discussion where we bring together the core findings emerged from our research.

\section{Combining Art, Science and Sound Research}

A dialogue between scientific research and artistic practice is inclined to generate multiple and often contrasting perspectives, rather than producing an agreement. Important insight can emerge from those contrasts, insight that would have been obscured by a unified and monodisciplinary approach. This is the methodological grounding of the present chapter. We do not strive for objective knowledge. Art, even in its practice-based forms, does not offer any provable truth but creates alternate visions and, sometimes, it is called upon to “contemplate the dark without drawing a resolutely positive lesson”, as Fuller and Goriunova put it \cite{Fuller2019}.

\subsection{Practice-based Research and Objective Knowledge}

The kind of art and science coupling we discuss here is not the use of scientific findings for the creation of an artwork, for that strategy requires an attentive process of translation, as stressed by Stengers~\cite{Stengers2000}, and – based on our subjective experience – it can easily feel constraining. As we will elaborate later when discussing our first collaborative project \emph{Corpus Nil}, valid insight or correct measurements achieved in a laboratory setting may not be useful when applied to a musical performance. The motivation lies in the experimental limitations of particular scientific methods found in HCI and other research fields combining natural science, design and engineering. Contrary to the social sciences, for instance, HCI experiments are often conducted in a laboratory context. In controlled  studies, every aspect of such studies is thoroughly directed by the experimenters. Whereas the protocols of such HCI experiments aim to ensure scientific rigour and reproducibility, those same protocols create the specific context wherein the collected data and related findings are meaningful. Outside of the specified context, data and findings may not be as meaningful, especially when applying them to an unpredictable and ever-changing situation such as a live artistic performance before a public audience.

For example, during a live performance, sensor measurements that have been validated through a particular experiment may very well vary according to a range of factors, including room temperature, magnetic interferences and the like. Those variables - contrary to a laboratory experiment situation - cannot be controlled. More importantly, the performer experiences her own body, a technological musical instrument and the context in subjective ways; during an artistic performance a performer is driven by instincts, desires and feelings that arise from the relation with the audience, other performers and a particular musical instrument. Artistic expression in musical performance emerges from the interaction of human actions and desires with the capabilities of a musical instrument.

This emphasises the need for, on one hand, a careful selection and interpretation of the methods of art and science collaboration,  and, on the other, a precise definition of the shared knowledge needed by artist and scientist to operate collaboratively; an entanglement that Roger Malina aptly calls “deep art-science coupling”~\cite{Malina2006}. Practice-based research represents a good framework for the exploration of art and science coupling, for it generates knowledge from action, self-reflection and empirical experimentation~\cite{candy2006practice,biggs2004learning}, methods that are equally fitting in HCI as in music and the performing arts. Crucially, we believe that the goal of art and science collaboration is not the production or re-staging of the limits of a normative type of science, but rather is about ``working outside current paradigms, taking conceptual risks'', following Malina, so as to mindfully merge disciplines towards new practices of experimentation. This can result in contrasts and disimilarities which must be considered and which, in fact, produce the richness of art and science practice.

\subsection{Artistic Intervention in Scientific Research}

As technology has become more ubiquitous, spreading from workplaces to  everyday life, the field of HCI has adopted new ways to investigate how we interact with such pervasive technology \cite{hutchinson2003technology,dourish2004action}. The so-called third-wave of HCI looks at notions such as experience, emotions, aesthetics and embodiment \cite{bodker2006second}. The field thus embraces methodological resources from a broader range of fields such as cognitive science, social science, science studies, or the Arts.
In this context, artistic intervention encourages research practitioners to look at their subject of study under an alternative “hypothesis”, through different facets. It is important to underline that, differently from practice-based research, artistic intervention in research emphasizes the idea of intervention: the artistic process \emph{acts} upon the research objectives. For example, artistic intervention in scientific research, and more specifically in HCI, can help address a problem (or study a phenomenon) in radically new ways \cite{benford2012uncomfortable}. Such as in the case of Fdili Alaoui \cite{fdili2019making} who, working between HCI and dance, questions the academic culture of HCI through her expertise in interactive dance performance.

In our research, we use artistic intervention to build and study interactions between people and intelligent systems. This methodological choice is motivated by two factors. On one hand, scientific research practice can feel normative and standardised. As researchers, we employ established methods to structure our investigations on a particular question, as well as to disseminate the outcomes of the research. While this standardisation has its pragmatic rationale (for instance, to help its dissemination), it can be detrimental to scientific research practice in many ways, such as limiting exploration or surprise. A detailed articulation of the pros and cons of it goes beyond the scope of this paper. Here we stress how artistic intervention in research can complement standard scientific methods in leaving more room for unexpected challenges and nuanced conceptual questioning. As an example, which will be developed in the following section, our study of gestural expressivity in musical performances has shifted the emphasis from characterizing gestural variations in kinematic space to characterizing variations in muscle coordination space.

On the other hand, the specific field of HCI has its own culture and shortcomings. While the field examines and generates the systems underpinning the technological world with which humans interact daily,  its political dimension  -- that is, the varying balances of power between who produces the technology, who designs it, who uses it, and what socio-cultural impact it may have -- is rarely addressed~\cite{irani2010postcolonial,bardzell2015immodest, keyes2019human}. By avoiding to explicitly address the politics of technological innovation, one risks to contribute to the creation of forms of technological solutionism. Technological solutionism refers to the idea that technology can solve any problem, including issues which may not exist or which ignore both socio-cultural and political contexts \cite{blythe2016anti}. We found that one method to prevent (at least partially) our research from entering a solutionist discourse is to invoke artistic intervention within the process of scientific research. Critical artistic intervention can explicitly provoke self-reflection about the involvement of a particular technology. Through an analytical and self-reflexive engagement, a critical artistic intervention can question the role of technology and its capacities of meaning-making, forcing artists and researchers to face the inner beliefs that ultimately motivated their choices.

In the following, we discuss how we applied the methods and insight described above to two separate research projects, which led to the creation of two artistic performances and scientific publications. This also serves to illustrate how our collaboration developed iteratively through the intertwining of practice-based research in the art field and artistic intervention in the scientific research.

\section{Machine Learning as a Tool for Musical Performance}

Our first project stemmed from a collaboration started in 2014. The outcomes of this particular project were a performance entitled \emph{Corpus Nil} (premiered in 2016 at ZKM, Center for Art and Media, Germany and still touring internationally) and academic publications \cite{caramiaux2015understanding, donnarumma2013muscular}.

\subsection{\emph{Corpus Nil}}
\begin{figure}[!ht]
    \centering
    \includegraphics[width=1\columnwidth]{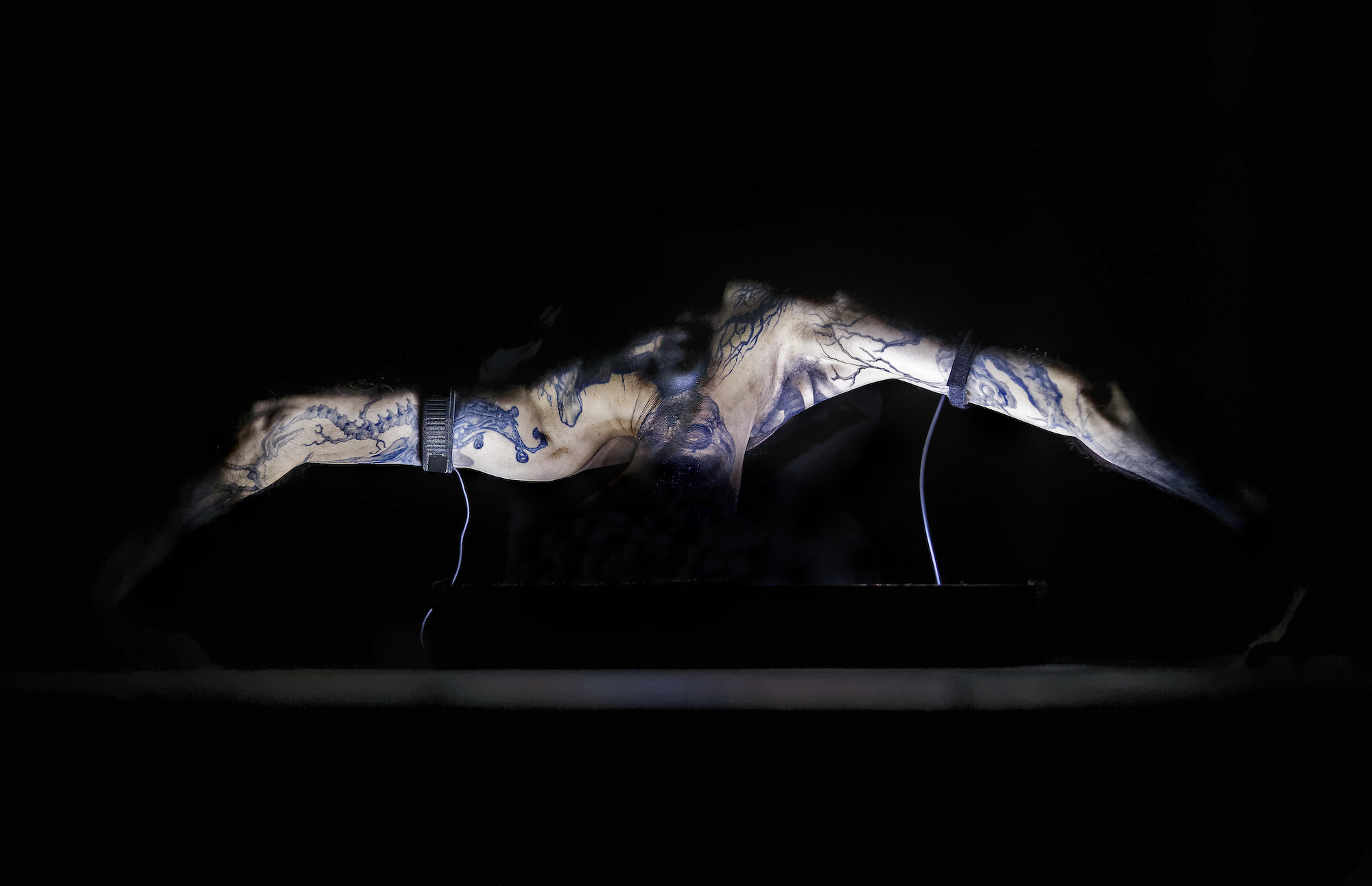}
    \caption{Picture from the performance \emph{Corpus Nil}. Photo courtesy of ONUK Fotografie. The piece evolves through different levels of muscular tension, torsion and articulations of the upper limbs. Aspects of the movements captured by means of two different muscle sensors placed on the arms.}
    \label{fig:Corpus}
\end{figure}

The piece is a twenty-minute dance and music performance exploring hybrid forms of identity and musicianship. It is an intense and ritualistic interaction between an autonomous musical instrument, a human body, and sound. Figure~\ref{fig:Corpus} depicts a picture from a live performance.  The theater space is completely dark. The player (Donnarumma in this case), whose body is partly naked and partly painted in black, performs a tense choreography which gradually morphs his body. Two types of wearable biosensors transmit data from his body to our machine learning-based software. Chip microphones capture sounds from muscles and internal organs (mechanomyogram or MMG) and electrodes capture muscle voltages (electromyogram or EMG). The software uses special filters to generate a description of the amplitude and frequencies of all sounds produced within the performer's body (between 1-40 Hz), as well as their variations over time. Then, it re-synthesises those sounds by orchestrating a feedback network of twenty digital oscillators. Because the choreography demands slow, subtle and iterative physical movements, the resulting music is equally slow and recursive, mutating across microtonal variations of a minimal set of pitches.

The instrument – here intended as a combination of our chosen hardware and software – analyses the collected data to learn the nuances of the performer's movement (muscular tension, gesture abruptness, rate of relaxation). Upon observing a certain nuance, it chooses whether to mute or activate particular oscillators, how to regulate volumes, phases, glissandos and multi-channel diffusion, and how to adjust feedback amounts within the network. The player cannot control the instrument, but can only learn how to affect it and be affected by it. The piece thus discards conventional performer-instrument relationships – founded on the performer's full control of the instrument – in favour of an unstable corporeal engagement between the two. Through the rhythm of sound, vibration and light, the performer's body and the instrument mutate, physically and conceptually, into something “other”; an unfamiliar creature defying common definition of the human.

The software used in \emph{Corpus Nil}, collaboratively created by the authors, consists of a set of algorithms linked to one another forming a recursive network. The algorithms have different tasks, such as biosignal analysis, learning of movement nuances (through linear regression and statistical methods), and sound resynthesis. A group of algorithms extracts a set of high-level features from the biosignals collected from the player's body (change rate, damping and spectral centroid). This feature set does not represent the movement per se, for it does not account for the shape of movement in space. Rather, it indicates specific traits, such as muscular force, abruptness of a contraction, and damping of the muscular tissues, that characterise the articulation of the movement.
The combination of analysis, learning and sound synthesis algorithms into a recursive network makes the instrument dynamic enough to provide highly unexpected responses to the player's movements. This, in turn, forces the performer to adapt the choreography “on the fly”, establishing thus a continuous and adaptive dialogue with the instrument.

\subsection{Scientific and artistic drives}

Here we present the scientific and artistic contexts, specific to each author, in which our collaboration on \emph{Corpus Nil} took place.

\paragraph{Gesture Expressivity and Interactive Machine Learning}

When we started the collaboration in 2014, my (first author) research focused on capturing movement expressivity to be creatively used in music performance. A hypothesis was that aspects of the physical expressivity of a performer could be observed by analysing intentional variations in the execution of a movement. A system which would capture such variations could then be used to drive sound synthesis engines. However capturing those variations proved not trivial. Our approach at the time, developed with colleagues at Ircam (Paris) and Goldsmiths College (London), relied on a machine learning-based algorithm – called Gesture Variation Follower (GVF) \cite{caramiaux2014adaptive} – which tracks, in real-time, continuous movement variations across space and time. In order to allow artists with radically different movement vocabularies and expressivity to use the system, this was designed to be rapidly calibrated.

This research belongs to a broader line of endeavour where machine learning algorithms are considered as tools to design expressive musical instruments and interactions~\cite{fiebrink2011human, caramiaux2013machine, gillies2016human, fiebrink2018machine, gillian2014gesture, franccoise2014probabilistic}. In this type of research, the significant advantage of machine learning is that it leverages advances in computation and data analysis to allow musicians to tackle fairly complex musical scenarios~\cite{fiebrink2018machine}. More precisely, instead of writing rules that govern the criteria of interaction between performed movements and resulting sound, one can provide the algorithmic system with demonstrations of those criteria, which will be automatically learned by the system \cite{franccoise2015motion}. This approach involves tight interactions between the performer and the ML algorithm, formalised under the discipline of Interactive Machine Learning (IML)~\cite{ware2001interactive,fails2003interactive,amershi2014power}.
Using machine learning as a design tool seemed a natural  approach when considering movement inputs to an interactive system. Writing rules that govern the analysis of movement inputs is clearly unfeasible when we consider that the system is expected to handle a wide scope of complex movements, mostly subjective to each performer. In addition, movement literacy is tacit, which means it cannot be easily formalised through words and lines of code of a programming language in a design context~\cite{hook2010transferring}. A machine learning-based approach where one can configure a system by demonstration seems a much more efficient approach~\cite{gillies2019understanding}. In addition, such a system can be more inclusive, meaning that it can be used by novices and people with diverse physical abilities, who may not have previous experience in movement analysis or computer science~\cite{katan2015using}.

\paragraph{Deconstructing a Performer's Body}
By the time our collaboration began, I (second author) had been working for four years in the field of gestural music as a researcher, composer and performer; in particular, I had been developing a musical performance practice known as \emph{biophysical music}~\cite{Donnarumma2017}. With this term, I refer to live music pieces based on a combination of physiological technology and markedly physical, gestural performance. The approach of biophysical music differs significantly from previous strands of biosignal-based musical performance~\cite{Knapp1990, Rosenboom1990, Tanaka1993} which are largely rooted in the use of physiological signal as a means of control over electronic or digital interfaces. The practice of biophysical music focuses, instead, on designing ways and tools through which the physical and physiological properties of a performer’s body are interlaced with the material and computational qualities of the electronic instruments, with varying degrees of mutual influence. Through this method, which I term human-machine \emph{configuration}~\cite{Donnarumma2017beyond}, musical expression arises from an intimate and, often, not fully predictable negotiation of human bodies, instruments, and programmatic musical ideas (an anthology of biophysical music projects can be viewed in~\cite{Donnarumma2015}).

After several years exploring potentialities and limits of gestural music research, I started questioning – in both my artistic and scientific processes –  the kind of human body at the center of conventional gestural music performances. My concern arouse from both my own work experience in the field and my participation to the HCI community through conferences and research exchanges, as well as from my study of feminist theory, in particular body theory~\cite{Blackman2012} and disability studies~\cite{Shildrick2002}. Soon, it became  clear to me that the kind of body shown in the performances of gestural music, and new musical instruments in general, was most times a heavily normalised body, that is, a body fitting particular criteria of normality. This triggered me to investigate how elements of gestural music performance and HCI could be exploited – in combination with research on movement and aesthetic – to create a different kind of performance practice; a practice that would account for the human body as an ever-changing, fluid entity with multiple potentials and varying forms of embodiment, instead of a static, controlled and controlling subject as established by our societal regime. Thus, drawing on the artistic work of performers such as K\={o} Murobushi~\cite{Murobushi} and Maria Donata D’Urso~\cite{Disorienta}, among others, I developed a particular methodology of movement research that emphasised the combined use of unconventional choreographic methods and intense somatic experimentation towards the physical manipulation and symbolic deconstruction of a performer’s body.

The movement experiments I conducted at the time were based on a gesture vocabulary composed of complex torsions, flexions, and contractions of shoulders, upper arms, and neck. No lower arm or hands gesture figured in the movement vocabulary I had designed for the piece (which then became \emph{Corpus Nil}); hands, lower arms and even the head were, in fact, hidden from view and rendered purposely useless. Because the literature and tools in regard to this particular mode of gestural interaction are scarce, we decided to explore muscle activity at a very low level -- through choreographic methods and resources from biomedical engineering. This, in turn, led us to investigate complementary sensor modality, with the goal to understand physical expression outside of a frame focused on ``control''.

\subsection{Development and Observations}

The initial seed of the collaboration was an intuition to investigate applications of interactive machine learning to the analysis of movement in a body-based musical performance. The performance in question, however, relied on the expression of bodily and motion qualities that were uncharted by previous work in the field. This spawned a research of muscle activity, the sensorimotor system and how, through different sensing methods, aspects of expressivity could be observed in human-computer interaction.

\paragraph{Understanding Movement Expressivity}
Initially, the analysis of movement expressivity as conveyed by the specific kind of muscle torsions, flexions and contractions described in the previous section seemed a good challenge for the Gesture Variation Follower algorithm (GVF)~\cite{caramiaux2014adaptive}. GVF was originally designed to track variations of movement trajectories in space and time, so we explored fundamental questions about movement expressivity at the level of the muscle activation.  What is the trajectory of muscle activity in this context? To what extent information on muscle temporal activity is meaningful to a performer, and how muscle temporal activity can be controlled?

We began by experimenting with mechanomyogram (MMG) sensors, capturing the mechanical activity of the muscle, built by Donnarumma in previous work~\cite{Donnarumma2011icmc}. We organised sessions of data capture and post-analysis. It became quickly clear that MMG data captured dynamic transitions between separate muscle contractions and the resulting trajectories calculated by GVF did not make sense. We added a second sensor based on electromyogram (EMG), which measures the electrical discharges activating the muscle. Here a new problem emerged, following Donnarumma’s feedback about the resulting movement-sound interaction: controlling EMG trajectories was not aesthetically convincing, for it could not allow a highly dynamic mode of interaction, one of our main musical goal. Therefore, we worked on a new way to analyse muscle biosignals. We developed a different tracking system where the tracked parameters were not linked to the movement trajectory but the to the parameters of an underlying second order dynamical system with damping (symbolically representing muscle dynamic as an oscillating string).

How to define gesture expressivity when the gestures in questions operate through a set of symbolic and aesthetic signs? What kind of expressivity emerges from a performer's body that is physically constrained by a given choreography? And how can a ML system actively and subtly support a semi-improvisational physical performance of music where the very notion of ``control'' is put into question? We felt that these questions were to be tackled through a strategy that combined scientific and artistic methods. Our research forked in two parallel streams of investigation. On one hand, we deepened our understanding of expressivity to take into account involuntary aspects of whole-body gestures. Using real-time analysis of complementary muscle biosignals, EMG and MMG, we began isolating a set of features that could describe relative levels of muscular effort without the need to define in advance the type or the timing of a gesture. This would support the open-ended and semi-improvised nature of the choreography that was being developed. On the other hand, we started exploring the design of a computational music system that would not be tied to conventional mapping techniques, but instead would reflect, through sound, levels of effort intensity, movement abruptness and degree of complexity of particular gestures – i.e., the amount of muscle groups activated by a given gesture.

We finally conducted a controlled experiment to study gestural expressivity from the point of view of motor control (i.e., the ability of users to control gesture variations), using both EMG and MMG interfaces~\cite{caramiaux2015understanding}. The main study brought together 12 participants for a session of 45 minutes each. In these sessions, participants were asked to perform a certain number of gestures (with or without tangible feedback) and then to vary one or more dimensions of expressivity. We showed that the participants consistently characterized dimensions of implicit expressivity (force, tension, etc.) and that the physiological interfaces used made it possible to describe these dimensions. Finally, we showed that participants could control these dimensions under certain conditions linked to constraints from the laws of motion.

\paragraph{Highlighting algorithm's limitation with IML}
GVF was designed to be quickly trained (or calibrated) by providing one example for each movement that the system has to recognise, and its capacity to track movement variation was meant to give to users a means to explore a movement space. However its use in a tight interaction loop with the performer’s body during experiments and rehearsals felt like a failure.
One major factor was that the system felt restrictive rather than extensive. The movement possibilities seemed to shrink rather than extend. In the typical IML workflow, the performer is engaged in a close interaction loop with the movement-based machine learning system. However, if the algorithm has low-capacity (failing to handle a wide range of input movements) with respect to the complexity of the input movement, the performer may ultimately adapt to the system limitations, hence constraining her own movement qualities. Consequently, this adaptation would make the interaction converge towards simplistic movements and interactions.

This phenomenon may have also been emphasized by the absence of clear explanations about the system's prediction behavior. Incorrect analysis of the GVF were hard to grasp by the performer (second author). Typically, the algorithm could have failed because the beginning of the movement is too different from the pre-recorded ones, but this information was not clearly fed back to the performer. As most temporal gesture recognition systems, GVF recognises a gesture (or movement) from a starting point until an end point. Establishing what defines the beginning of a movement and its ending point during the artistic performance was not trivial and it could definitely not be handled by the algorithm.  Moreover, the system was designed to handle continuous and rather slow variations, whereas Donnarumma’s muscle activity during the choreography presented types of dynamics that could not be tracked by the system. We needed a more flexible approach and thus we developed a new tracking system. This was designed to estimate in real-time the parameters of a physical model that mimicked the performer's muscular activity, as mentioned above. In doing so, we were not tracking gesture trajectories but movement ``regimes''. Thus, this method allowed us to explore alternative representations of movement using the input data at hand.

\section{Artificial Intelligence as Actor in Performance}

In 2019, we initiated a second collaboration, which actively involves a group of 13 people, including artists working with new media, biological materials, and performance, as well as designers, AI and neurorobotics scientists. The project outcome is a dance-theater production entitled \emph{Humane Methods}. The first version of the performance premiered at Romaeuropa Festival and, at the time of writing, the project is currently ongoing. As for the previous collaborative project, we begin by describing the piece and then situating it within distinct scientific and artistic contexts.

\subsection{\emph{Humane Methods}}

\begin{figure}[!ht]
    \centering
    \includegraphics[width=1\columnwidth]{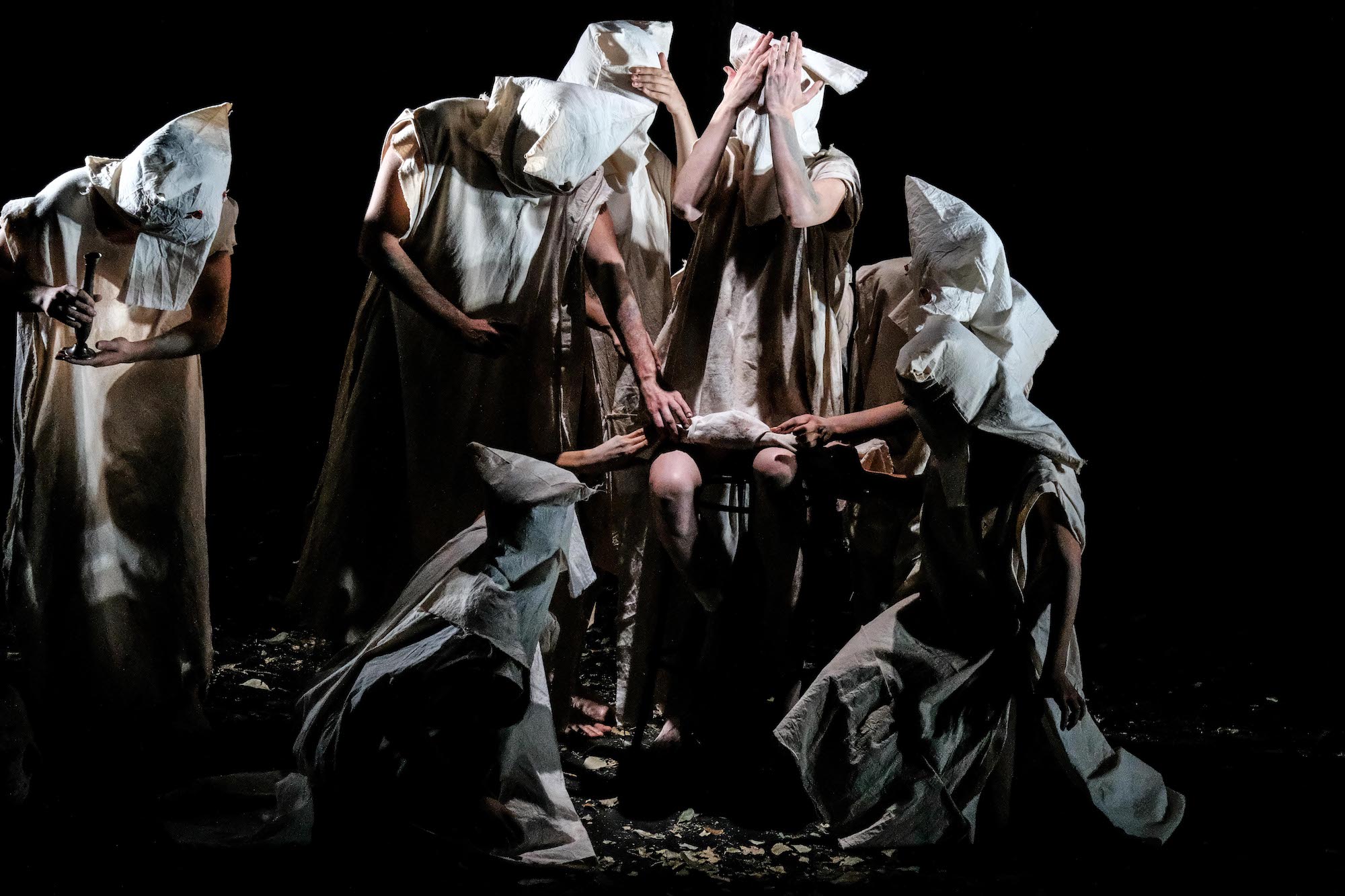}
    \caption{Photo courtesy of Giada Spera. Picture from a live performance of Humane Methods at Romaeuropa Festival. Portrayed is one of the iterations of the ritual at the core of the performance. It is possible to observe in detail the lighting generated by the AI.}
    \label{fig:humane}
\end{figure}

\emph{Humane Methods} is a dance-theater production exploring the multilayered nature of today’s violence. The project departs from the assumption that – through the physical and psychological brutalization of people and non-human beings – the combination of digital technologies and capitalistic urge has driven the natural ecosystem towards impending destruction. The project then aims to dissect the violence of algorithmic societies, where power structures, knowledge creation and normative criteria become means of manipulation. Being an evening-length production, \emph{Humane Methods} is composed of a multitude of elements, including robotic prosthetic limbs driven by their own neural networks, dead plant specimen and fungi growing on costumes, uncompleted or destroyed architectural structures, chimeric creatures, experiments in trauma, actions as rituals. For the sake of clarity, here we focus on the technological aspects of the piece and how they relate to the techniques and implications of AI. Figure \ref{fig:humane} shows a picture from the performance.

In a landscape of ruined nature and technological relics, a community of nine semi-human figures lives a primal and stern existence. They perform a ritual of prayer in a loop, but with every repetition of the prayer, something in their interpersonal relations or in the world surrounding changes. As variations accumulate, the praying loop is drastically altered and the storyline branches out into a constellation of stories, of hints, of untold events, playing out various degrees of intolerance, isolation and exploitation. Throughout the action, a deep reinforcement learning algorithm tries to learn a meaningless sequence of ten digits.

Our motivation in choosing to work with an AI algorithm performing meaningless calculations lies in a conceptual shift from conventional uses of deep learning techniques in music and performance.
What interests us is not the capacity of the algorithm to reach its target, but rather the ways in which the inner (obsessive) logic of this type of computation can be made perceivable at an aesthetic and sensorial level to both performers and audience. To that end, we designed an audiovisual system that links the actions of the AI algorithm to synaesthetic patterns of music and light, so that the algorithm “speaks”, as it were, of its actions and choices through sound and light. The result is a dogged, hypnotic stream of sound and light that inundate the venue: the calculations of the AI are turned into perceivable material, auditory and visual pulsating patterns that literally mark the choreographic actions on stage, as well as the spectators’ experience of the piece.

\subsection{Scientific and Artistic Drives}

\paragraph{Human - AI Interaction}
ML and AI algorithms have known an exponential development in the past decades. Several important works have shown the capacity of learning-based algorithms to outperform humans in specific cognitive tasks such as playing the game Go \cite{silver2016mastering}, recognising images \cite{krizhevsky2012imagenet} or understanding natural language \cite{hinton2012deep}.

Simultaneously, the fields of HCI, Humanities, and Social Sciences have started to take seriously the study of the socio-cultural impact of AI and to propose design guidelines or good practices. Several works have recently been published along this line. Among them, some aim to propose a set of guidelines for designers of AI-powered interactive systems \cite{amershi2019guidelines}. In this context, AI is seen as a tool to improve users’ services and generally empower humans \cite{cai2019hello}. As an extension, AI can be seen as a material that can be appropriated by designers to improve user experience in AI-based applications \cite{dove2017ux}. However, most of these attempts have highlighted the inherent challenges of this approach due to the difficulty of grasping what AI can and cannot do \cite{yangre}. In my work (first author), I am exploring how the computational learning mechanisms themselves can become interactive in order to foster exploration and human learning, as we recently explored in the specific domain of sound design~\cite{scurto2019designing}.

On the other hand, recent works have investigated the complex consequences of the inherent biases in AI-powered applications and their impact when used in health care, education or culture \cite{angwin2016machine,kulesz2018culture,crawford2019excavating}. This last branch of research is a necessary undertaking that I am currently pushing in my recent research interest, which looks at the socio-cultural impact of technology. Therefore, at a scientific level, the collaboration presented in this section was motivated by an intention to investigate how people understands AI technology and which are the founding beliefs supporting  particular forms of such understanding.

\paragraph{Motivations and Artistic Needs}
The project \emph{Humane Methods} is a collaborative endeavour that involves, aside from the collaboration between myself (second author) and Caramiaux, a shared authorship between Margherita Pevere \cite{Pevere}, myself and video artist Andrea Familari. In particular, the artistic concept for the piece was jointly created by Pevere and myself, and then furthered through a close collaboration with Caramiaux. As artists working respectively with biotechnology and AI technology, the first question that Pevere and I asked ourselves was how to situate our artistic practices within the current historical period, one characterised by a generally intolerant and polarised socio-political forces, converging with a serious, technologically-driven disruption of the natural environment. This question led our efforts towards an analysis of the multifaceted nature of human violence, as it is exerted between humans as well as it is enforced by humans on the lives of non-human entities and the natural environment.

Turning to the main topic at hand -- AI technology and its use in collaborative art and science projects -- we chose to deploy AI algorithms in an unconventional way. Instead of using AI to aid human performers in the creation of sound – as we had done for \emph{Corpus Nil} – or using an AI generative capacity to create a  musical automata, we chose to make tangible the ``brute force'' computing mechanism of AI algorithms. In other words, we chose to highlight how many of the most common learning algorithms being presently developed attempt to reach their targets using a mechanism that is obsessive and raw.

\subsection{Development and Observations}

The collaboration developed through personal exchanges on the general theme of technology and society and particularly on how technology induces violence or empathy on people, to other people and to their environment. We wanted to create an algorithmic behavior that obsessively learned something and which could autonomously explore different ways to reach its goal. The first trials, in terms of scenography, music, light and algorithms happened in June 2019.

\paragraph{Machine Behaviour}
The algorithm designed and implemented for \emph{Humane Methods} is based on deep reinforcement learning, which means that the algorithm explores a high-dimensional continuous space and receives positive rewards when approaching the goal. These rewards are used to exploit certain moves within this space that would help the algorithm reach the goal. In \emph{Humane Methods}, the algorithm’s target is an arbitrary set of parameter values which -- purposely -- holds no meaning. The reward is given by its distance to the target. Its actions are moves in the parametric space. As the algorithm moves in the parametric space, it outputs a list of 10 values – indicating the degree of proximity of its calculations in respect to its target on an integer scale between 1 and 3. Each of the 10 values is then mapped to a musical pattern – which are predefined arpeggios of acoustic piano samples – and a light pulsation pattern. Sonic and light patterns follow relentlessly one after the other, creating the perception of an audiovisual entity morphing continually through a multitude of syncopated variations. This dynamic morphing of sound and light is made more complex by a parallel method linking the algorithm degree of proximity to the target to, on one hand, the loudness of each single musical pattern, and on the other, to the brightness intensity of physical analog lights. The closer a value is to the corresponding target, the quieter is the volume of the related musical pattern and the dimmer is the brightness of the corresponding light pattern.

The algorithm moves within the space are entirely inferred by the learning process. This process consequently makes apparent to the spectators the behaviour of the algorithm. The algorithm becomes thus an actor, a performer on its own. Because neither the human performers nor the technical team behind the piece can control the computation of the algorithm, the latter is the sole entity responsible for the audiovisual dramaturgy of the piece. Interestingly, this role taken by the algorithm made us personify it: we began referring to it by a name, or discussing its contribution to the piece as an actor on its own. Journalists have also asked questions along this line, after seeing the show. This has been all the more striking for us as we were very careful not to anthropomorphize AI, a process that we find counter-creative for it typically impedes a precise understanding of technology by imposing a human worldview onto something which is not human.

\paragraph{Controlling Artificial Intelligence and Being Controlled by It}
Following our first experiments with the algorithm and, thus, observing the modalities of its interaction with music, light and choreography, it became more interesting to us \emph{how} the algorithm learned rather than what it learned. We tested several strategies to map, more or less directly, the behaviour of the algorithm (the how) to the choreography executed by the human performers – discarding, in the process, any information regarding what the algorithm learned. In doing so, we soon faced a different challenge: how to find a balance between the control of the AI over the piece's narrative and the influence of the performers  over the AI's learning process. We explored different strategies. One was to have the algorithm dictate the start and end points of each choreographic action. This, however, felt narratively constraining and aesthetically poor. Another attempt required the AI to describe the movement qualities with which each loop was to be performed by the human actors. This proved cumbersome, for a reliable, real-time communication between the AI and the performers was too complex to design.

At this point we changed our approach. Instead of mapping the AI's behaviour to the actions on stage, we thought of implementing the opposite strategy, that is, to map the actions on stage to the behaviour of the AI.  Thus the choreography was re-conceived to mimic the temporal structure of the AI's learning process.  One single action (a hybrid form of gestural prayer) is repeated in a loop, but with each repetition comes a more or less drastic variation.  This loop structure mirrors the episodic learning of the algorithm: the algorithm learns through episodes -- each can take a few minutes -- and each episode adds variations to the next iteration of learning. This solution meant to limit the explicit interaction between actors’ performance and AI learning process. The interaction between the two elements became thus implicitly intertwined: at conceptual and aesthetic levels the performers' action and the algorithm's learning process were connected, but this connection was not defined by explicit mapping of control signals. While we successfully performed the piece before a public audience in this format, we consider the issue still open. In our view, the piece would have a stronger impact if the performers' actions would be linked \emph{both} implicitly and explicitly to the learning process of the AI.

\paragraph{Audience Perception of Artificial Intelligence}
In June 2019, we had the opportunity to collect subjective feedback from the audience during an open rehearsal at the Centre des Arts Enghien les Bains (CDA), a co-producer of \emph{Humane Methods} together with Romaeuropa Festival, Italy. Our encounter with the audience in France was important, for it allowed us to open our aesthetics and research to a general public, most of whom had little or no familiarity with the ongoing project or our previous works. Here, we focus on the feedback given by the audience about their perception of the AI, as they observed it during the excerpts of the piece we performed for them. It is important to note that, before the performance, we purposely avoided informing the audience about the presence of an algorithm driving lights and music.

The two most significant comments of the audience were that some spectators did not realize that music and light were driven by an AI algorithm, and that another spectator suggested us to create more explicit interactions between the algorithm and the performers on stage. These comments are perhaps unsurprising if we consider the issue of implicit/explicit interaction discussed above. Nevertheless, they confirmed our early intuition about the challenges of making an AI perceivable to a public audience. Our algorithm manifests itself only indirectly, through lights and music and it lacks a physical embodiment. This was a choice we made in order to reinforce our critical stand on the impact of AI in society; the AIs regulating social media, welfare and warfare are pervasive and ubiquitous, while simultaneously invisible and implicit, integrated seamlessly as they are in institutional structures and interpersonal lives. Our AI was designed with this in mind, and is therefore omnipresent and unobservable.

However, concept aside, on an actual theater stage it is indeed a challenge to manifest the AI's presence and expressivity while completely avoiding a physical representation of it (and potentially fall into naive anthropomorphism).
Thus, while we still believe the concept we explored is meaningful to the overall piece, the audience feedback made us realize that, in order to be perceived as a “real” presence, the AI in \emph{Humane Methods} needs to be embodied and its agency has to be manifested, in a way or another. This is still ongoing research, thus we  can only speculate here, but we believe this issue opens a fascinating challenge regarding  the fragile balance between “AI in control” versus “humans in control” elaborated in the previous section.

\section{Discussion}

This chapter focused on possible methodological approaches to the use of ML and AI in the context of music and performance creation. We have offered an autoethnographic perspective on a research and creation process spanning five years. Our aim was to open up our process and proffer insight into one possible way of coupling art and science in the application of AI and ML. In closing this chapter, we discuss three main aspects of the research. The first relates to how we approached AI and ML in relation to music and sound creation. The second concerns a shift in terminology that accompanied, and became manifested through, our projects. The third and final one addresses the hybrid methodological approach we developed progressively and contextually to each artwork.

\subsection{Artificial Intelligence and Music}
Many music-related fields are currently facing important changes due to the intervention of machine learning and artificial intelligence technology in music processing.
These changes occur at different levels: creation, consumption, production and diffusion \cite{caramiaux2019ai}. While AI technologies offer increasingly broad options to researchers and artists alike, with the present contribution we stress the importance of acknowledging the historicity of these technologies – where and how they emerged and developed – as well as the increasingly complex implications of their interaction with multiple layers of humans' societies.
As we have demonstrated through the review and discussion of our specific collaborative projects, the use of ML and AI in music-related research does not have to be constrained by a paradigm of control – of humans over algorithms, of humans over musical forms, or of humans over other humans. Rather, by nurturing in-depth and long-term art and science collaborations it is possible to define new modalities of interaction across ML, scientific research and artistic creation.

At a musical level, both \emph{Corpus Nil} and \emph{Humane Methods} foster improvisational and unconventional types of interaction between algorithms, human bodies and musical forms. What should be emphasised, however, is that these kinds of musical interactions emerged from the combination of particular scientific and artistic drives which, in turn, evolved out of specific socio-cultural contexts. Artistic intervention into scientific research was coupled with practice-based research in ways that ramified our thoughts and strategies in multiple directions. Our own idea of what computational music can be was heavily affected by the particular paths we chose throughout the past five years of collaborative work. Whereas this allowed the germination of nuanced and technically rigorous interaction systems, it also made the development of satisfying outcomes slower than it could have been, had either of chosen to work alone. Thus, the same combination of practices and approaches which affords us with the capacity to deeply couple art and science, is a factor that impedes fast turnarounds and requires mid-large timescales in order to reveal artistically and scientifically valid outcomes.

\subsection{From Machine Learning to Artificial Intelligence}

Our work witnessed a change in terminology that has been occurring since the deep learning breakthrough in 2012, that is, the shift from ``machine learning'' to ``artificial intelligence''. Machine learning used to -- and, to some extent, still does -- refer to the technological underpinnings that allow computational systems to ``learn'' from data, where learning means extracting structural components from a dataset. The term artificial intelligence is however more ambiguous: it was originally coined in the 1950's to designate a symbolic approach for modeling human cognition, but nowadays what is called AI mostly relies on vector-space generated by deep neural networks (which is not a symbolic approach)~\cite{cardon2018neurons}. The term has its advantage to communicate about the technology and to create research and business incentives. Then, we also believe that the term imposed itself following the recent capacities of machine learning techniques. When AlphaGo~\cite{silver2016mastering} won against the world’s best Go player, the idiom ``machine learning'' did not seem comprehensive enough to designate the technology that made it possible, its computational capacities, or the new prospect that such technology had opened up.

While, today, the HCI community often uses the term Artificial Intelligence, in our own research we preferably adopt the term Machine Learning to designate the type of algorithms we develop and use. Nevertheless, in our view, the adoption of the term AI triggers interesting questions into the concepts that are most commonly associated with the term. Speaking of AI triggers a specific and ambiguous imaginary that is worth understanding in detail in order to make better use (or no use at all) of this technology as practitioners and researchers in HCI.

The dialectical shift can be traced in our own work. Our first project, \emph{Corpus Nil}, relied on a low-capacity movement recognition system. In this case, we were certain that Machine Learning was the right designation because the algorithm did not manifest a distinctive behavior. Rather, its role was to recognize and learn given aspects of a performer's physical expressivity, so as to emphasize elements of human corporeality. On the contrary, the second project, \emph{Humane Methods}, involved a radically different algorithm chosen not for its capacity to achieve a certain goal, but rather for the particular computational behavior driving its learning mechanism. In this case, the term Machine Learning felt constraining, for it precluded an understanding of the algorithm as a socio-cultural phenomenon, beyond its computational capabilities. The algorithm had almost become a black box, interpretable only by observing its choices and subsequent actions. In this sense, it became observable as an actor.

\subsection{Hybrid Methodology}

By observing the varied methods we deployed in the research presented in this chapter, what comes to the fore is a hybrid methodology, an adaptive range of methods which evolved according to the nature and context of the particular problem at hand. This is illustrated by how, in \emph{Corpus Nil} and \emph{Humane Methods}, learning algorithms were used in two fundamentally different ways, and how, in each case, the coupling of the respective needs of art and science led to unconventional and effective results.

In \emph{Corpus Nil}, we used an algorithmic system to design expressive, body-based musical interactions between a human performer and algorithms. We set out to explore Machine Learning algorithms as a \emph{tool}, therefore our focus was less on the algorithm itself and more on what results the algorithm could achieve during the performance. The particular choreography created for the piece revealed limits of available computational approaches for the design of interactions with sound, and led us to find an ad hoc method based on physiological sensing. This method -- triggered by artistic intervention into scientific research -- yielded a new approach to the analysis of gesture expressivity, one based on complimentary aspects of muscle physiological data. This allowed us to capture dimensions of expressivity that greatly differ from those obtained through computer vision interfaces, which are the most commonly used in the literature.

In \emph{Humane Methods}, we designed a computational framework which eventually became an actor, an independent entity directing sound and light in a dance-theatre performance according to its own internal functioning. Contrarily to \emph{Corpus Nil}, at the core of this piece was an investigation of AI as a cultural and social \emph{concept}, therefore we were less concerned with the results the algorithm could achieve and more preoccupied with how to manifest the functioning of the algorithm itself and to exhibit its behavior. On one hand, this shift of focus proved challenging because it forced our research to confront a broad set of (scientific, artistic and popular) expectations surrounding the notion of AI. On the other hand, though, the challenge has opened before us a vast field of investigation, which extends across HCI, performing art and into the socio-political impact of AI. We feel that the prospect for a further entanglement of science, art and politics is, today, perhaps more urgent than ever.

\section{Acknowledgements}

We would like to thank Fr\'ed\'eric Bevilacqua (IRCAM, Sorbonne Universit\'e), Sarah Fdili Alaoui and Margherita Pevere for their helpful comments on the paper. This research was partly supported by the European Research Council under the European Union's Seventh Framework Programme (FP/2007-2013) / ERC Grant Agreement n. 283771; by the CNRS under the programme PEPS / project INTACT; and by La Diagonale Paris-Saclay, under the programme Phares 2019 / project Humane Methods.

\small

\bibliographystyle{abbrv}

\end{document}